\documentclass[iop]{emulateapj} 
\usepackage{natbib}
\usepackage{graphicx}
\usepackage{color}

\graphicspath{{./}}

\newcommand{\Msun}{\rm M_{\odot}}

\newcommand{\MBH}{\rm M_{\rm{BH}}}

\newcommand{\Mstar}{\rm M_{*}}
\newcommand{\Mgal}{\rm M_{\rm{gal}}}
\newcommand{\Mhost}{\rm M_{\rm{host}}}
\newcommand{\Msph}{\rm M_{\rm{sph}}}
\newcommand{\mbhsigma}{\MBH\rm{-}\sigma}

\newcommand{\mbhsph}{\rm{\MBH\rm{-}\Msph}}
\newcommand{\mbhhost}{\rm{\MBH\rm{-}\Mhost}}
\newcommand{\mbhgal}{\rm{\MBH\rm{-}\Mgal}}

\newcommand{\Mpc}{\rm {Mpc}}

\newcommand{\lya}{\ifmmode {\rm Ly}\alpha \else Ly$\alpha$\fi}
\def\msunyr{\ifmmode M_{\odot} {\rm yr}^{-1} \else M$_{\odot}$ yr$^{-1}$\fi}

\newcommand{\na}{\rm {New Astronomy}}

\begin{document}                          

\title{Scaling Relations Between Low-mass Black Holes And their Host Galaxies}    

\author
{
Qirong Zhu\altaffilmark{1,2},
Yuexing Li\altaffilmark{1,2},
Sydney Sherman\altaffilmark{1,2}
}

\affil{$^{1}$Department of Astronomy \& Astrophysics, The Pennsylvania State University, 
525 Davey Lab, University Park, PA 16802, USA}

\affil{$^{2}$Institute for Gravitation and the Cosmos, The Pennsylvania State University, University Park, PA 16802, USA}

\email{qxz125@psu.edu} 

\begin{abstract}

It is well established that supermassive black holes in nearby elliptical galaxies correlate tightly with the kinematic property ($\mbhsigma$ correlation) and stellar mass ($\mbhhost$ correlation) of their host spheroids. However, it is not clear what the relations would be at the low-mass end, and how they evolve. Here, we investigate these relations in low-mass systems ($\MBH \sim \rm{10^{6}- 10^{8}}\, \Msun$) using the Aquila Simulation, a high-resolution cosmological hydrodynamic simulation which follows the formation and evolution of stars and black holes in a Milky Way-size galaxy and its substructures. We find a number of interesting results on the origin and evolution of the scaling relations in these systems: (1) there is a strong redshift evolution in the $\mbhsigma$ relation, but a much weaker one in the $\mbhhost$ relation; (2) there is a close link between the $\mbhsigma$ relation and the dynamical state of the system -- the galaxies that fall on the observed correlation appear to have reached virial equilibrium. (3) the star formation and black hole growth are  self-regulated in galaxies -- the ratio between black hole accretion rate and star formation rate remains nearly constant in a wide redshift span $z = 0-6$. These findings suggest that the observed correlations have different origins: the $\mbhsigma$ relation may be the result of virial equilibrium, while the $\mbhhost$ relation may the result of self-regulated star formation and black hole growth in galaxies. 

\end{abstract}

 
\section{INTRODUCTION}

A major development in observational astrophysics in recent years is the discovery that most, if not all, nearby elliptical galaxies host a supermassive black hole (SMBH) at their center \citep{Kormendy1995}, and that the masses of the SMBHs correlate tightly with the global properties of the spheroid components of their hosts, such as the stellar velocity dispersion (the $\mbhsigma$ correlation, e.g., \citealt{Ferrarese2000, Gebhardt2000, Tremaine2002, Gultekin2009A, Graham2011}), and the stellar masses (the $\mbhsph$ correlation, e.g., \citealt{Magorrian1998, Marconi2003, Haring2004}). These correlations suggest that the formation and evolution of SMBHs and their host galaxies are closely linked (e.g., \citealt{Haehnelt2000}).

The origin of these scaling relations, however, remains a hot debate. A number of different models have been proposed to explain the correlations, such as  feedback from active galactic nuclei (AGN, e.g., \citealt{Silk1998, Ciotti2007}), gas competition between star formation and BH accretion (e.g., \citealt{Li2007A, Jahnke2011}), galaxy mergers (e.g., \citealt{Peng2007}), and a combination of merger and AGN feedback (e.g., \citealt{Li2007B, Hopkins2006A, Hopkins2007B, Hopkins2009A, Hopkins2009B}). 

In order to unravel the origin of the correlations, recent observational efforts have focused on their evolution in hope to identify the crucial physical processes in galaxy formation which give rise to the relations (e.g., \citealt{Shields2003, Treu2004, Walter2004, Shields2006, McLure2006, Peng2006, Woo2006, Salviander2007, Treu2007, Woo2008, Jahnke2009, Bennert2010, Decarli2010, Merloni2010, Cisternas2011, Bennert2011}). These surveys suggest that there is either weak or no evolution in the $\mbhsigma$ relation up to $z \sim 2$, but there is a strong redshift evolution in the $\mbhsph$ relation. However, it was also suggested that the relations between $\MBH$ and total host-galaxy luminosity or stellar mass, $\mbhhost$, may not be evolving, or at least not as rapidly as the $\mbhsph$ relation (e.g., \citealt{Jahnke2009, Bennert2010, Merloni2010, Cisternas2011, Bennert2011}). 

These results suggest that the $\mbhsph$ and $\mbhsigma$ relations have different origins. More importantly, the difference between $\mbhsph$ and $\mbhhost$ provides a crucial clue on the origin and evolution of the mass scaling relation, because $\mbhsph$ is closely connected to the formation of bulges, which is believed to result from major mergers \citep{Barnes1992B, Hernquist1992, Hernquist1993, Hopkins2006A}, and thus depends strongly on redshift and the merging history of a galaxy \citep{Hopkins2010A, Hopkins2010B}. On the other hand, $\mbhhost$ may simply reflect the growth of stars and BHs in a galaxy regardless its type. Therefore, the $\mbhhost$ may represent a more fundamental property of galaxies. In this work, we adopt the latter as it is also easy to measure from the simulations.

Theoretically, the evolution of the BH -- host relations has been studied using numerical simulations (e.g., \citealt{Robertson2006, DiMatteo2008, Johansson2009, Booth2011}) and semi-analytical models (e.g., \citealt{Hopkins2006B, Hopkins2009B, Malbon2007, Somerville2008, Somerville2009, Lamastra2010, Kisaka2010, Zhang2012}). However, these models differ in their predictions for the evolution of the scaling relations. In particular, \cite{Robertson2006} employed a set of idealized merger simulations at various redshifts and found that the slope of the $\mbhsigma$ relation remains roughly constant at redshifts $z=0 - 6$, but the normalization shows a weak redshift dependence. \cite{DiMatteo2008} performed the first direct cosmological hydrodynamic simulation with BHs down to redshift $z=1$, and found a weak redshift evolution in the normalization and slope of both correlations. \cite{Booth2011} also used a cosmological simulation but a different BH model and found that, while they reproduced the local correlations and the observed evolution of $\mbhhost$ for massive galaxies, they predicted an evolution in the $\mbhsigma$ which contradicts with current observations. 

These studies have painted a confusing picture of the BH -- host correlations. The confusion may come from the different samples used, and the different evolutionary stages of these galaxies. The galaxy samples used to derive the original correlations are dominated by local massive ellipticals and inactive galaxies. Recently, it has been suggested that the correlations vary with galaxy type and mass. \cite{Gultekin2009A} showed different scaling relations for ellipticals and for spirals, and the latter have a larger intrinsic scatter in the $\mbhsigma$ relation than the ellipticals.  \cite{Kormendy2011} suggested  that the BH masses do not correlate with galaxy pseudobuldges or disks. Moreover, \cite{ Lauer2007B} found that, at the very high mass end, the BH masses derived from $\mbhsigma$ differ those from $\mbhsph$ by nearly one order of magnitude. 

At the low mass end  ($\MBH \lesssim 10^8 \Msun$), most of the observed galaxies are active galactic nuclei (AGN). It was reported that, compared to inactive galaxies, AGNs show a larger dispersion in the $\mbhsigma$ relation, and that the slope differs from that of inactive galaxies or massive ellipticals (e.g., \citealt{Greene2008, Greene2010, Kuo2011, Bennert2011, Xiao2011}). More interestingly, these authors found that while their samples show a large deviation from the ``classical'' $\mbhsigma$ relation, they seem to follow the $\mbhsph$ relation. Recently,  \cite{Decarli2012} compiled a sample of 26 quasars in the low-mass end ($\MBH \sim \rm{10^{7}- 10^{9}}\, \Msun$), and found that the $\mbhsph$ relation is consistent with that of the quiescent galaxies. 

In order to understand the origin of the scaling relations, it is important to follow the formation and evolution of BHs and their host galaxies, in particular at the low-mass end, because these are the building blocks of the massive ones at the present day. Based on our current understanding of structure formation in the cold dark matter (CDM) cosmology, the central BHs may follow the same hierarchical assembly of the galaxies as well. Black holes in the mass range of $10^{5-8}\, \Msun$ thus serve as a bridge between the SMBHs and their seeds, thus would provide crucial information on the growth of BHs. 

In this work, we study the BH -- host relations and their evolution in low mass systems using a high-resolution cosmological hydrodynamic simulation which includes important physics of dark matter, gas dynamics, star formation, black hole growth, and feedback processes. The simulation used the Aquila initial condition \citep{Scannapieco2012}, which was constructed to produce a Milky Way-size halo and its sub-structures (referred to as ``Aquila Simulation'' hereafter). We also explore the origin of the relations by comparing model predictions with observations. The paper is organized as follows: in \S 2, we describe our models and computational methods. We present the evolution of the relations in \S 3, models for  their origins in \S 4, and summarize in \S 5.

\section{The Aquila Simulation}
%\label{sec:method}

\begin{figure*}
\begin{center}
\includegraphics[width=2.2in]{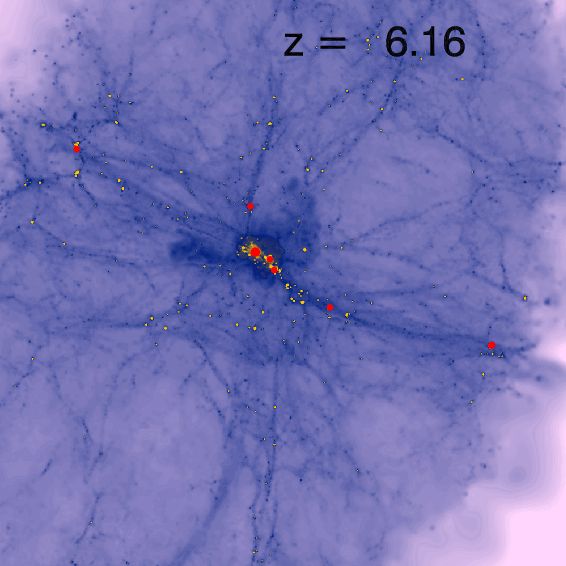}
\includegraphics[width=2.2in]{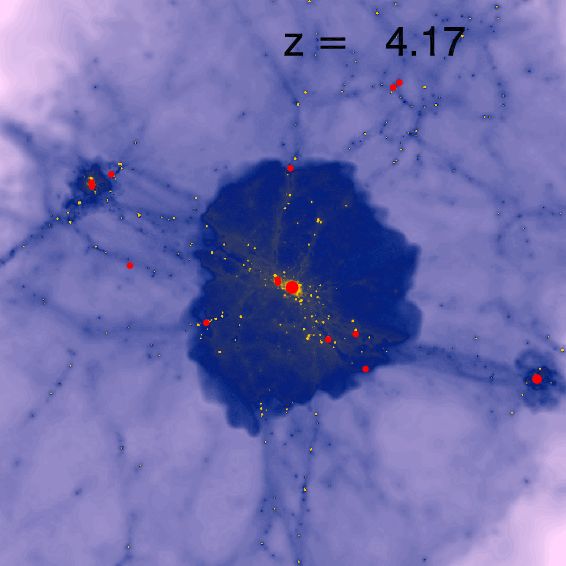}
\includegraphics[width=2.2in]{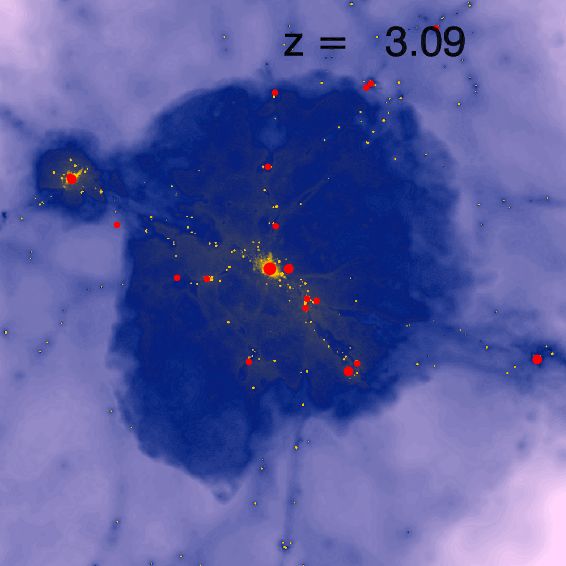} \\
\includegraphics[width=2.2in]{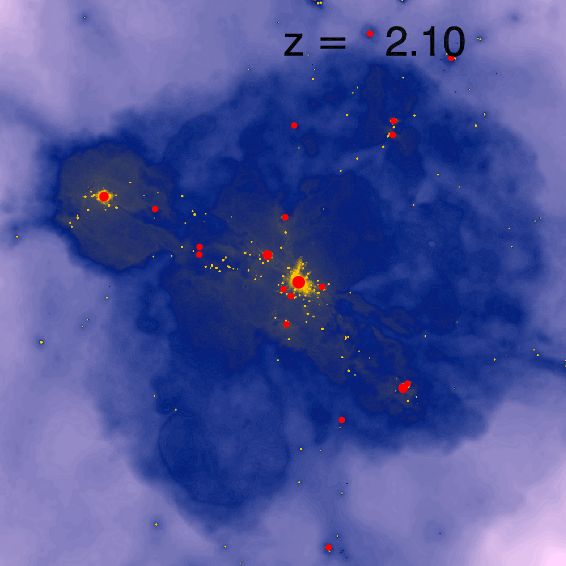}
\includegraphics[width=2.2in]{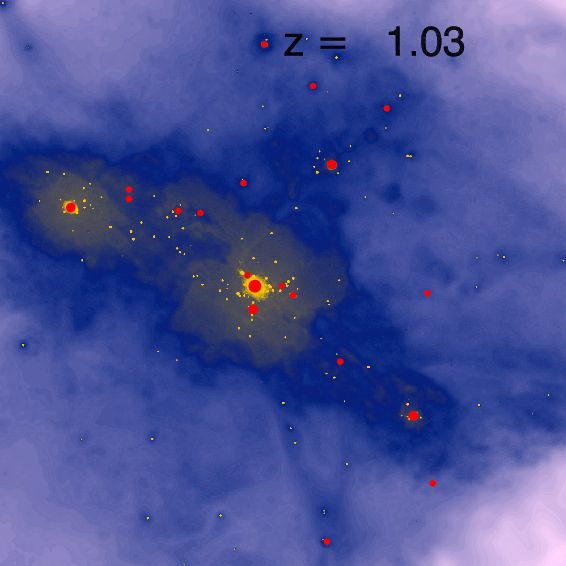}
\includegraphics[width=2.2in]{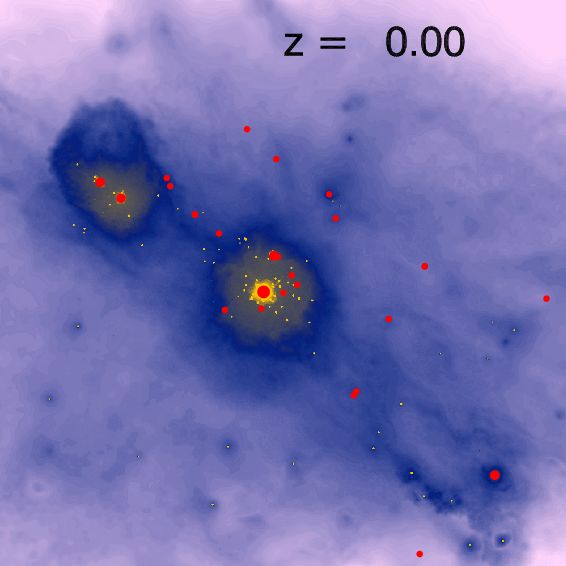}
\caption{Evolution of the BHs and galaxies from the Aquila Simulation. The images show the projected density of both gas (color-coded by
  temperature: blue indicates cold gas, brown indicates hot, tenuous  gas) and stars (represented by the bright yellow color). The black holes are represented by the red dots, the size of which is proportional to the BH mass. Feedback from both supernovae and accreting BHs creates hot bubbles around the galaxy centers. The box size is 10 $h^{-1} $Mpc in comoving coordinates.
}
\label{fig:img}
\end{center}
\end{figure*}

In order to achieve high resolutions in a cosmological simulation to study the BH -- galaxy correlations at different cosmic time, we carried out the Aquila Simulation, which follows the formation and evolution of a Milky Way-size galaxy and its substructures \citep{Wadepuhl2011, Scannapieco2012}. The initial condition is originally from the Aquarius Project \citep{Springel2008}, which produced the largest ever particle simulation of a Milky Way-sized dark matter halo. The hydrodynamical initial condition is reconstructed from the original collisionless one by splitting each original particle into a dark matter and gas particle pair  \citep{Wadepuhl2011}. 

The Aquila Simulation includes dark matter, gas dynamics, star formation, black hole growth, and feedback processes. Star formation is modeled in a multi-phase ISM, with a rate that follows the Schmidt-Kennicutt Law (\citealt{Schmidt1959, Kennicutt1998}). Feedback from supernovae includes both thermal and kinetic forms. Thermal feedback is captured through a multi-phase model of the ISM by an effective equation of state for star-forming gas, and the kinetic feedback is modeled as a galactic wind based on \cite{Springel2003}. We adopt a constant wind velocity of $v_{\rm wind} = 484~\rm{km~s^{-1}}$, a mass-loss rate that is twice of the star formation rate, and an energy efficiency of unity such that the wind carries $100\%$ of the supernova energy. The wind direction is anisotropical, preferentially perpendicular to the galactic disk. This wind model causes an outflow of gas, transporting energy, matter and metals out of the galactic disk in proportion to the star formation rate.

The model of black hole growth and feedback follows that of  \cite{Springel2005A} and \cite{DiMatteo2005}, where black holes are represented by collisionless ``sink'' particles that interact gravitationally with other components and accrete gas from their surroundings. The accretion rate is estimated from the local gas density and sound speed using a spherical Bondi-Hoyle \citep{Bondi1952, Bondi1944, Hoyle1941} model that is limited by the Eddington rate. Feedback from black hole accretion is modeled as thermal energy, $\sim 5\%$ of the radiation, injected into surrounding gas isotropically. This feedback scheme self-regulates the growth of the black hole and has been demonstrated to successfully reproduce many observed properties of local elliptical galaxies (e.g,, \citealt{Springel2005A, Springel2005B, DiMatteo2005, Hopkins2006A}) and the most distant quasars at $z \sim 6$ \citep{Li2007B}. Mergers of black holes can happen once two black holes are close enough and their relative speed is less than the local sound speed. In the simulation,  the black hole seeding scheme follows that of previous work \citep{DiMatteo2008, Sijacki2009, DiMatteo2012}:  a seed black hole of mass ${\MBH} = 10^{5}~ h^{-1} \Msun$ is planted in the gravitational potential minimum of each new halo identified by the friends-of-friends (FOF) group finding algorithm with a total mass greater than $10^{10}~ h^{-1} \Msun$. 

The computation was performed using the parallel, N-body/SPH code GADGET-3, which is an improved version of that described in \cite{Springel2001, Springel2005C}. For the computation of gravitational forces, the code uses the ``TreePM'' method \citep{Xu1995} that combines a ``tree'' algorithm \citep{Barnes1986} for short-range forces and a Fourier transform particle-mesh method \citep{Hockney1981} for long-range forces. GADGET implements the entropy-conserving formulation of SPH \citep{Springel2002} with adaptive particle smoothing, as in \cite{Hernquist1989}. Radiative cooling and heating processes are calculated assuming collisional ionization equilibrium \citep{Katz1996, Dave1999}, and the UV background model of \cite{Haardt1996} is used, which describes a spatial uniform UV background leading to reionization roughly at $z \approx 6$ in the simulation. 

The whole simulation falls in a periodic box of $100~h^{-1} \Mpc$ on each side with a zoom-in region of a size $5\times 5\times 5~h^{-3}\Mpc^{3}$. The spatial resolution is $\sim 250~h^{-1}$ pc in the zoom-in region. The mass resolution of this zoom-in region is $1.97 \times 10^{5}~ h^{-1} \Msun$ for dark matter particles, $\sim 1.875  \times  10^{4}~ h^{-1} \Msun$ for gas and star particles. The cosmological parameters used in the simulation are $\Omega_{m }= 0.25$, $\Omega_{\Lambda} = 0.75$, $\sigma_{8} = 0.9$ and $h=0.73$, consistent with the five-year results of the WMAP \citep{Komatsu2009}. The simulation evolves from $z = 127$ to $z = 0$.

In the simulation, each snapshot is processed by an on-flying FOF algorithm which tests the dark matter linking length as if they are less than 20\% of their mean spacing. Gas and star particles are then linked to the nearest dark matter particle. A substructure detection algorithm SUBFIND (an extended version of \citealt{Dolag2009} for gas and star particle) is then applied to each group by calculating the local density and searching for locally overdense region. Such an overdense region will be marked as a substructure of the parent group. Throughout this work, a galaxy is defined as the group returned by the SUBFIND, which includes dark matter halo, gas, stars, and black holes. We only select galaxies from the high-resolution, zoom-in region.  

Figure~\ref{fig:img} shows the evolution of the BHs and galaxies from $z \sim 6.2$ to $z=0$ from the Aquila Simulation. The gas follows the distribution of dark matter in filamentary structures, and stars form in high density regions along the filaments. The most massive galaxy in the zoom-in region resides in the intersection of the filaments, the highest density peak in the simulated volume where gas concentrates in the deep potential well. The BHs form in these massive halos, and they grow through gas accretion and mergers following the hierarchical buildup of their host galaxies. The most massive BH resides  in the main halo at the present day, and there are about a dozen of BHs in the mass range of $10^{5} - 10^{8} \Msun$.

\section{The BH -- Host Relations and Their Evolutions}
\label{sec:results}

\subsection{The $\mbhsigma$ Correlation}

\begin{figure}
\begin{center}
\includegraphics[width=3.5in]{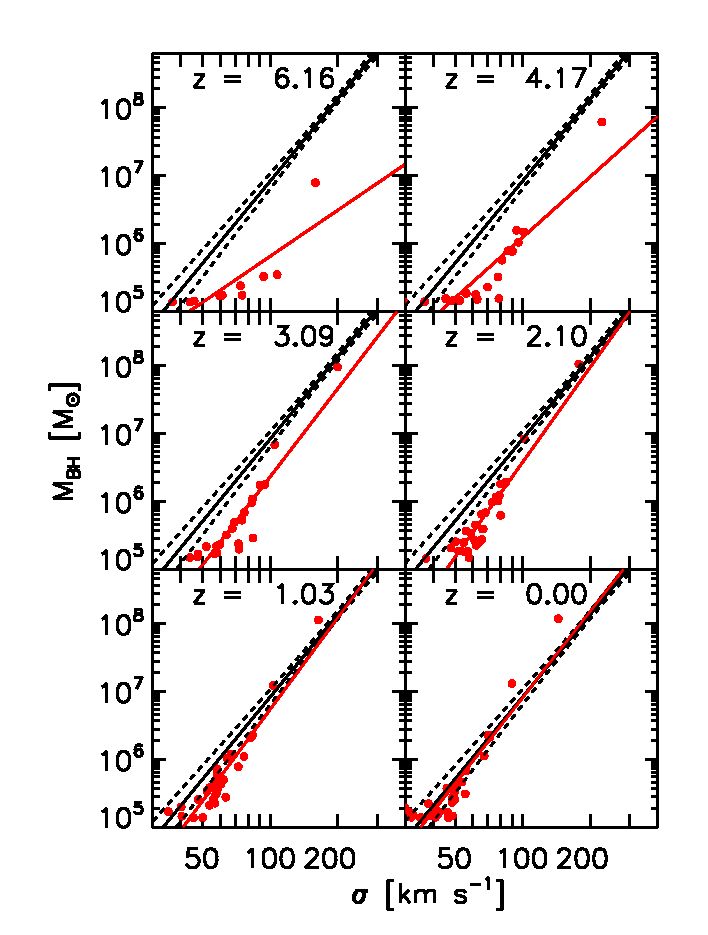}
\caption{\label{fig:msigma} The $\mbhsigma$ relation at different redshifts from the Aquila Simulation, in comparison with the local observation. The red dots represent the simulation data, the red solid curve is the fitting of our sample using a power-law formula, while the black solid curve is the best fit of the nearby galaxies by \cite{Tremaine2002}, and the dashed lines indicate the range of the fitting. }
\end{center}
\end{figure}

From the Aquila Simulation, we select a sample of galaxies with a stellar mass of $\Mstar \ge 10^{8} \Msun$, which contains over 5000 star particles. This criterion enables robust measurement of the galaxy properties. Most of the selected galaxies are dwarfs, and have actively accreting BHs. For each galaxy, the BH mass is computed directly in the simulation, and we calculate the projected half-mass effective radius $R_{e}$, and the line-of-sight mass-wieghted stellar velocity dispersion $\sigma$ within $R_{e}$. This projection procedure is performed for over 100 random line-of-sights. 

The resulting $\mbhsigma$ relation is shown in Figure~\ref{fig:msigma} at six representative redshifts, in comparison with the observed $\mbhsigma$ relation of the nearby galaxies by \cite{Tremaine2002}. We fit our data at each redshift with the same power-law formula as that in \cite{Tremaine2002}:
\begin{equation}
\log \frac{\MBH}{\Msun} = \alpha \log(\frac{\sigma}{\rm{200\, km\, s^{-1}}}) + \beta \,
\end{equation}
\noindent where $\alpha$ and $\beta$ represent the slope and the normalization of the relation, respectively. The best fit of the local $\mbhsigma$ relation from \cite{Tremaine2002} gives a slope of $\alpha=4.02\pm0.32$, and a normalization of $\beta=8.13\pm0.06$.

From Figure~\ref{fig:msigma}, there is a clear trend of evolution of the $\mbhsigma$ relation in both normalization and slope. The modeled galaxies systematically lie below the local $\mbhsigma$ relation at high redshift. A correlation emerges quickly after the BHs are formed, as high as $z = 4$, which is also the fast growth phase of BHs. As the galaxies evolve to low redshift, they converge to the observed local relation quickly. By $z=0$, some dwarfs eventually reach the line defined by local massive ellipticals, but the overall slope is noticeable steeper. The steeper slope at the low-mass end was also seen in simulations of \cite{DiMatteo2008}, and in low-mass AGN observations of \cite{Greene2010} from precise BH mass measurements using megamaser (note, however, \citealt{Xiao2011} reported a shallower slope with a large sample of AGNs.). The change in the $\mbhsigma$ relation at low mass probably indicates a different mode of BH growth in these objects compared with more massive galaxies with classical bulges \cite{Greene2008}. 

\begin{figure}
\begin{center}
\includegraphics[width=3.5in]{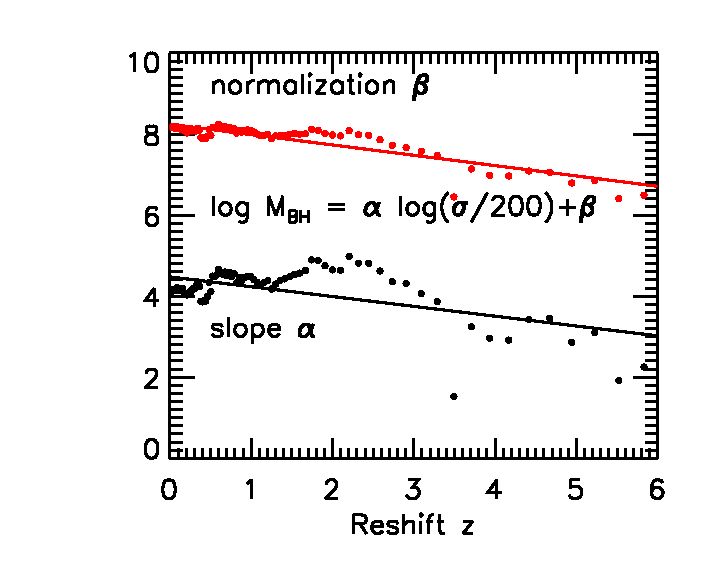}
\caption{\label{fig:msigma-ev} The evolution of the parameters of the $\mbhsigma$ relation, slope $\alpha$ (in black) and normalization $\beta$ (in red), with redshift in the simulation. The solid lines are least-squares fittings of the data.}
\end{center}  
\end{figure}

The evolution of the $\mbhsigma$ relation, in form of slope $\alpha$ and normalization $\beta$, is shown in Figure~\ref{fig:msigma-ev}. The scattering in both slope and normalization becomes larger at higher redshift, due to smaller galaxy number and increasing difficulty in the measurement of $\sigma$. Nevertheless, both slope and normalization show strong evolution with redshift out to $z=6$, as illustrated by the fitting lines: 
\begin{equation} 
\alpha = 4.38 - 0.21z, \beta = 8.07 - 0.20z 
\end{equation}

The evolution of the normalization of the $\mbhsigma$ relation from our simulation is consistent with the result of \cite{DiMatteo2008} from cosmological simulations, which also showed $\beta \propto - 0.20z$, and that of \cite{Robertson2006} from merger simulations, $\beta \propto - 0.186z$ by fixing the slope to 4.0, but shallower than that given by \cite{Booth2011}, $\beta \propto -0.32z$. Both \cite{Robertson2006}  and \cite{DiMatteo2008} suggested that the expected velocity dispersion for a given stellar mass is larger at higher redshift. As for the evolution of the slope, \cite{Robertson2006} did not see a monotonic trend possibly due to the limited number of redshifts in that study. Instead they put an upper limit on the absolute value of the evolution of slope, $ |\phi|< 0.3$. \cite{DiMatteo2008} showed that  the slope becomes steeper with decreasing redshift, similar to the trend we find, except that they have a much larger slope at $z \sim 3 - 4$ since the massive BHs ($> 10^8\, \Msun$) fall above the local correlation. The steep ``tilt'' at $z \sim 3 - 4$ in their work may be caused by the rapid growth of these massive BHs which gained their masses rapidly through mild super-Eddington accretion rates during this period. In fact, a similar trend is also present in Figure~\ref{fig:msigma-ev} at $z \sim 3 - 1$ when the most massive black holes in our simulation assembled most of their masses. 

In observation, the $\mbhsigma$ relation has been studied by using the width of some emission lines such as OIII or H$_{\alpha}$ as a proxy for stellar velocity dispersion \citep{Nelson1996, Gu2009}. These studies suggest that this scaling relation either does not evolve (e.g., \citealt{Shields2003}), or does so weakly, with BHs more massive than inferred from the local relation by a factor of a few at $z \sim 1$ (e.g., \citealt{Woo2008, Gu2009}). In Figure~\ref{fig:msigma-ev}, the evolution of the $\mbhsigma$ relation at $z \lesssim 1$ is much weaker than that indicated by the above fittings. At $z=0$, our result is within the observed range of low-mass BH systems \citep{Greene2010, Xiao2011}.

\subsection{The $\mbhhost$ Correlation}

\begin{figure}
\begin{center}
\includegraphics[width=3.5in]{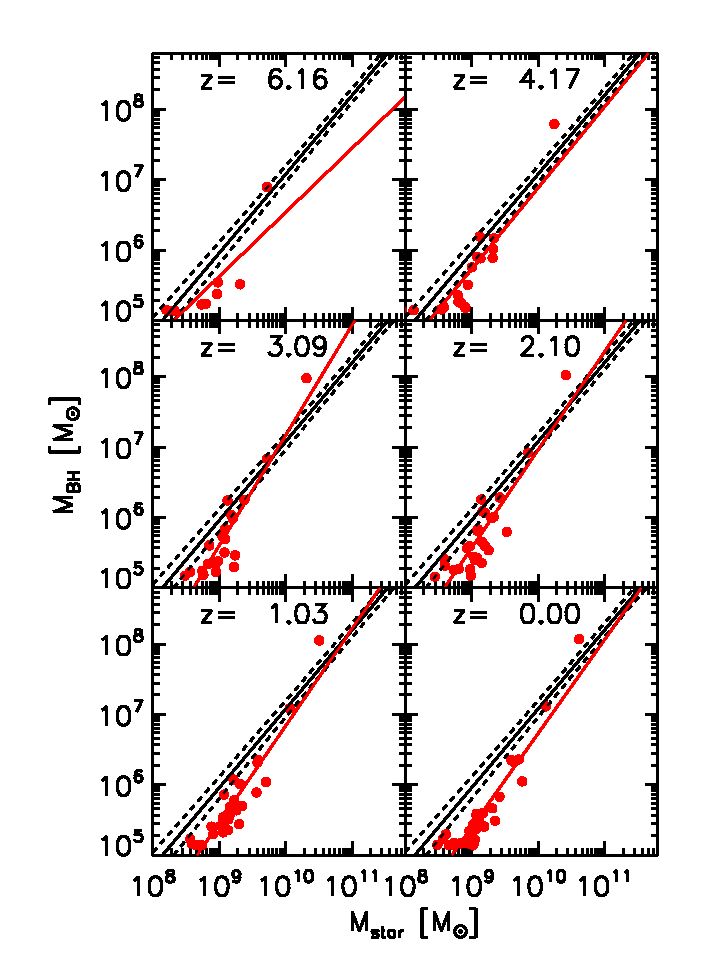}
\caption{\label{fig:mstar} The $\mbhhost$ relation at different redshifts from the Aquila Simulation, in comparison with the local observations. The red dots represent the simulation data, the red solid curve is the fitting of our sample using a power-law formula, while the black solid curve is the best fit of the nearby galaxies by  \cite{Haring2004}, and the dashed lines indicate the range of their fitting.}
\end{center}
\end{figure}

\begin{figure}
\begin{center}
\includegraphics[width=3.5in]{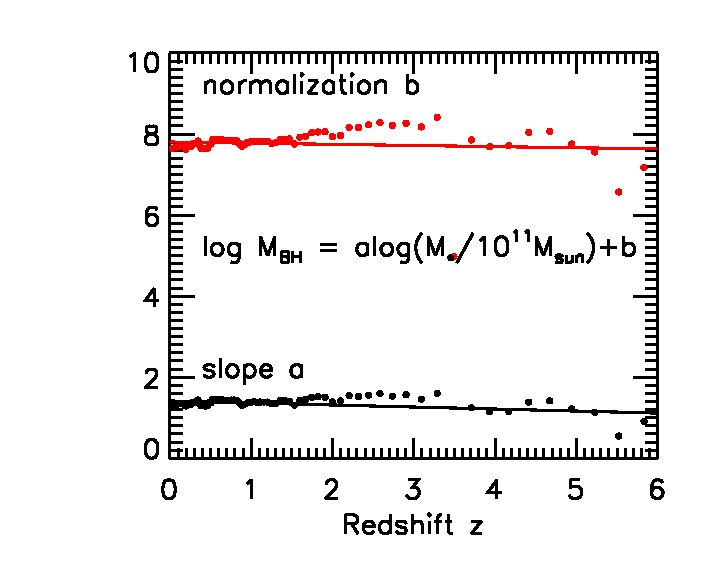}
\caption{\label{fig:mstar-ev} The evolution of the parameters of the $\mbhhost$ relation, slope $a$ (in black) and normalization $b$ (in red), with redshift in the simulation. The solid lines are least-squares fittings of the data.}
\end{center}
\end{figure}

For the same galaxy sample as in the previous section, we measure the masses of the hosts. Currently, we do not have an effective procedure to decompose the galaxy from a cosmological simulation into bulge and disk components, thus we are using half of the galaxy stellar mass as a proxy of bulge mass. The resulting $\mbhhost$ relation is shown in Figure ~\ref{fig:mstar}, in comparison with observations. We fit our data at each redshift with the same power-law formula as in \cite{Haring2004} : 
\begin{equation}
\log \frac{\MBH}{\Msun} = a \log(\frac{M_{*}}{\rm{10^{11} \Msun}}) + b 
\end{equation}
\noindent where $a$ and $b$ represent the slope and the normalization of the relation, respectively. The best fit of the local $\mbhhost$ relation from 
 \cite{Haring2004} gives a slope of $a=1.12\pm0.06$ and a normalization of $b=8.20\pm0.10$.

Compared to $\mbhsigma$ relation in Figure ~\ref{fig:msigma}, the $\mbhhost$ relation shows a weaker evolution with redshift. Most of the BHs in our simulation have masses $\MBH < 10^7\, \Msun$, and they lie below the local correlation, while the more massive ones $\MBH \gtrsim 10^7\, \Msun$, lies above the line, which in broad agreement with simulations of \cite{DiMatteo2008}, and  observations which have a  larger sample than ours, and they show a larger scatter above and below the local relation (e.g., \citealt{Greene2010, Bennert2011, Xiao2011}). 
 
The evolution of the $\mbhhost$ relation, in form of slope $a$ and normalization $b$, is shown in Figure~\ref{fig:mstar-ev}. The scattering in both parameters  becomes larger at higher redshift, due to smaller galaxy count, but on average, both slope and normalization do not show strong evolution with redshift out to $z=6$, as illustrated by the fitting lines. If we just focus on $z \lesssim 2$, there is no noticeable evolution in either the slope or the normalization. 
 
\begin{equation} 
a = 1.38 - 0.04z,  b = 7.64 + 0.02z 
\end{equation}

The modest evolution in the $\mbhhost$ relation from our simulation agrees well with the result from \cite{DiMatteo2008} in the same mass range. It appears that the BHs in massive galaxies ($M_{*}  \gtrsim 10^{10} M_{\odot}$ grow more quickly than the stellar mass \citep{DiMatteo2008, Booth2011}. This may be due to the super-Eddington accretion \citep{DiMatteo2008} and the increased efficiency early on \citep{Booth2011} in these models. \cite{Johansson2009} argued that SMBHs are less likely to develop before their parent bulges as they found that over massive BHs were not evolving to the local correlations from their merger simulations. However, \cite{Naab2009} suggested that minor mergers could be a solution to the increase of stellar masses. In our simulation, most of the BHs are low mass, and they accrete at sub-Eddington rates most of the time, which differs significantly from the massive ones in the previous studies. 

In observation, the evolution of the mass scaling relation has been studied by a number of groups (e.g., \citealt{Shields2003, Treu2004, Peng2006, Woo2006, Salviander2007, Treu2007, Woo2008, Jahnke2009, Bennert2010, Decarli2010, Merloni2010, Cisternas2011, Bennert2011}). In particular, it was recently suggested that the ratio between BH masses $\MBH$ and the stellar masses of the host spheroids $\Msph$ evolves strongly with time out to $z \sim 2$: $\MBH / \Msph \propto (1+z)^{1.96}$ \citep{Bennert2011}. However, most of the studies derived the bulge stellar mass from the luminosity of the spheroid. While this can be done robustly for local ellipticals, the uncertainty becomes larger at higher redshift, because the stellar populations in bulges are not well understood at $z \gtrsim 1$, it is more difficult to distinguish between the host and the bulge component, and bulge may not have formed in some high redshift galaxies or AGNs. In fact, when considering the total luminosity or the stellar mass of the host galaxy, it was found that the $\mbhhost$ has a much weaker evolution, if any, than the $\mbhsph$ (e.g., \citealt{Jahnke2009, Bennert2010, Merloni2010, Cisternas2011, Bennert2011}), consistent with our finding in Figure~\ref{fig:mstar-ev}.

Studies of quasar host galaxies indicate that $\mbhhost$ relation may be present as early as $z \approx 4$ \citep{Peng2006}, and that evidence for evolution of this relation is not conclusive because of the following three reasons: (1)  BH mass calibration has a factor of 2 systematic uncertainty;  (2)  bulge mass is not directly measured; and (3) selection bias. \cite{Decarli2010} studied 96 quasars and their host galaxies and found an evolution in $\mbhgal$ from $z = 3$ to the present day. However, the possibility that these host galaxies may be biased towards lower mass galaxies at high redshift is present, as pointed out by\citep{Lauer2007A}. Such a bias would lead to  a more massive black hole for a give luminosity by nearly one order of magnitude. On the other hand, a limited sample of obscured AGNs from \cite{Sarria2010} suggests that $\mbhhost$ relation still holds at high redshift. 

The difference between $\Msph$ and $\Mhost$ provides an important clue on the origin and evolution of the scaling relations, because the local $\mbhsigma$ and $\mbhsph$ are determined by the formation of bulges, which are believed from major mergers (e.g., \citealt{Hernquist1992, Hernquist1993, Hopkins2006A}). Overall, the evolution of the $\mbhsigma$ reflects the change of structural properties,  while $\mbhhost$ simply reflects the growth history of a galaxy and its central BH, regardless its type and or activity.

\section{Origin of the Scaling Relations}

The origin of the BH scaling relations has been an open question. Our simulation shows that there is a remarkable evolution with redshift in the $\mbhsigma$ relation, but little or no evolution in the $\mbhhost$ relation. The different evolutions of the scaling relations suggest that they may have different origins. 

\subsection{$\mbhhost$ Relation as a Result of  Self-regulated Growth in Galaxies }

\begin{figure}
\begin{center}
\includegraphics[width=3.5in]{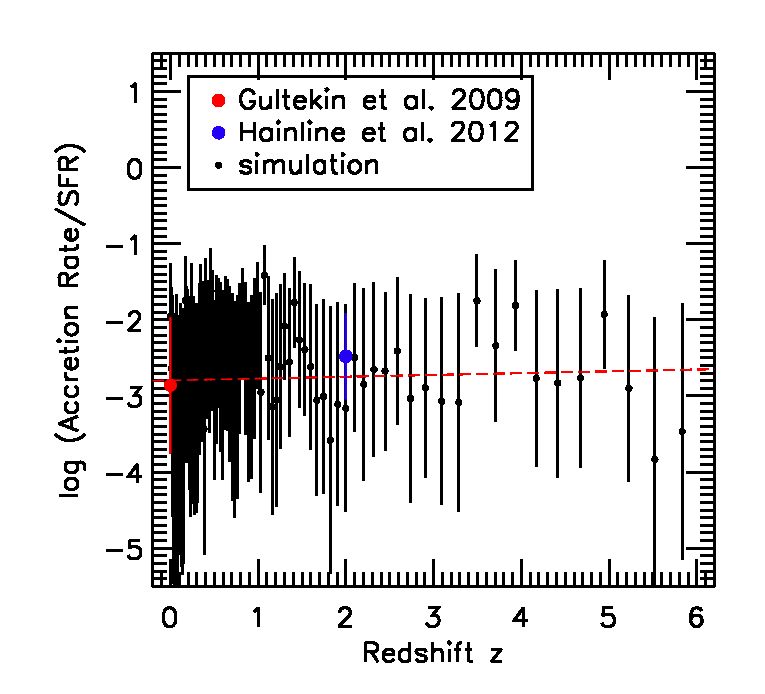}
\caption{\label{fig:growth} Ratio of BH accretion rate to star formation rate of our galaxy sample at different redshift from the Aquila Simulation. The black error bar indicates the range of the ratio at a given time, while the black dot represents the mean value. The red and blue dots represent observational data based on \cite{Gultekin2009B} and \cite{Hainline2012}, respectively. The red, dashed line is the least-squares fitting to the simulation data.}
\end{center}
\end{figure}

In order to understand the mass correlation, we turn to the growth history of the galaxies in our sample. Our simulation includes a number of physical processes of baryonic matter, including gasdynamics, cooling, star formation, BH accretion, and feedback from both SNe and BHs. Dark matter halo quickly forms and significant amount of mass is gathered through merger, leading to the reconfiguration of the gravitational potential and relaxation of DM particles. Gas falls into forming gravitational potential well and is shocked to high temperature. A considerable fraction of gas cools, loses its angular momentum, pile up in the center of the halo and stellar component builds up a short period of time, as well as the central existing BH. Feedback from SNe and AGNs heats up gas to hot phase, compensate the cold gas inflow and regulate both star formation and BH growth. The BHs grow until self-regulated, when its feedback energy can sufficiently unbind in-falling gas and halt accretion as well as star formation. The difference between an isolated merger and a cosmological simulation is that a constant gas supply from IGM is present in a cosmological simulation, which will exhibit as a cooling inflow to the galaxy in the early stage.

As a result of the interplay among these physical processes, the star formation and BH growth in galaxies appear to be self-regulated and balanced. Figure~\ref{fig:growth} shows the ratio of BH accretion rate to star formation rate of each galaxy at different redshift in the simulation, in comparison with two available observational data points at $z=0$ and $z \sim 2$, respectively. The local galaxy sample is based on X-ray luminosity from \cite{Gultekin2009B}, $H_{\alpha}$ luminosity from \cite{Ho2003} and infrared luminosity from \cite{Melendez2008}. The SFR is calculated from $\rm H_{\alpha}$ luminosity or far infrared luminosity using the conversion relations by \cite{Kennicutt1998}. The accretion rate is estimated with the same method as in \cite{Mullaney2012}. We convert X-ray luminosity to bolometric luminosity with a constant factor of 22.4, and assume a radiation efficiency of 0.1 to calculate the accretion rate. The total number of the galaxies in this sample is 12. The high redshift AGN sample from \cite{Hainline2012} includes 11 galaxies with both SFR and AGN bolometric luminosity. We calculate the accretion rate using the same method above. 

As shown in Figure~\ref{fig:growth}, although there is a large dispersion in the data due to variation in the BH accretion rate and star formation rate, the least-squares fitting to the simulation data (the red, dashed line) shows $\rm log (\dot{M}_{\rm BH}/SFR) = -2.754 - 0.02z $. The fitting indicates nearly negligible evolution in the ratio over redshifts $z=0 - 6$, and it lies close to the available observational data. More interestingly, this growth ratio, $\rm{\dot{M}_{BH}/SFR}=-2.754$, is very close to the mass ratio of the observed $\mbhhost$ relation, $\MBH/\Mhost \simeq (1-2) \times 10^{-3}$ (e.g., \citealt{Haring2004}).   

This result suggests that BHs have grown in-step with their host galaxies since at least $z \sim 6$, that the BH growth and star formation in galaxies are self-regulated, probably owing to feedback and availability of gas supply, and that the ratio of the two remains constant regardless the triggering mechanism. The fact that the ratio maintains constant at $\rm \dot{M}_{\rm BH}/SFR=-2.754$ naturally gives rise to the observed $\mbhhost$ relation in the local universe. 

Recently, \cite{Mullaney2012} studied the growth of SMBH and stars in a large sample of galaxies in the redshift range $0.5 < z < 2.5$. They found that the average SMBH accretion rate follows remarkably similar trend with stellar mass and redshift as the average SFR of their host galaxies, and that the ratio $\rm \dot{M}_{\rm BH}/SFR \simeq 10^{-3}$, similar to our finding in this work. Furthermore, \cite{Shi2009} studied infrared properties of 57 SDSS type -1 quasars at z$\sim$1 and found that the average ratio of $\rm \dot{M}_{\rm BH}/SFR$ in quasar hosts shows little evolution with redshift. These observations give strong support to our hypothesis that the observed $\mbhhost$ relation is a result of self-regulated and balanced growth in galaxies. 

Our model predicts that the $\mbhhost$ relation is a global, fundamental property of galaxies, it does not depend on galaxy type or mass,, and it does not evolve with redshift. However, it does not rule out strong evolution in the relation with {\it{spheroid}}, because the $\mbhsph$ depends on the formation of bulges, which is believed to be triggered by major mergers of galaxies, therefore, properties of spheroids depend sensitively on environment and redshift  which determine the merging history of a galaxy.

\subsection{$\mbhsigma$ Relation as a Result of Virial Equilibrium} 

\begin{figure}
\begin{center}
\includegraphics[width=3.5in, angle =0]{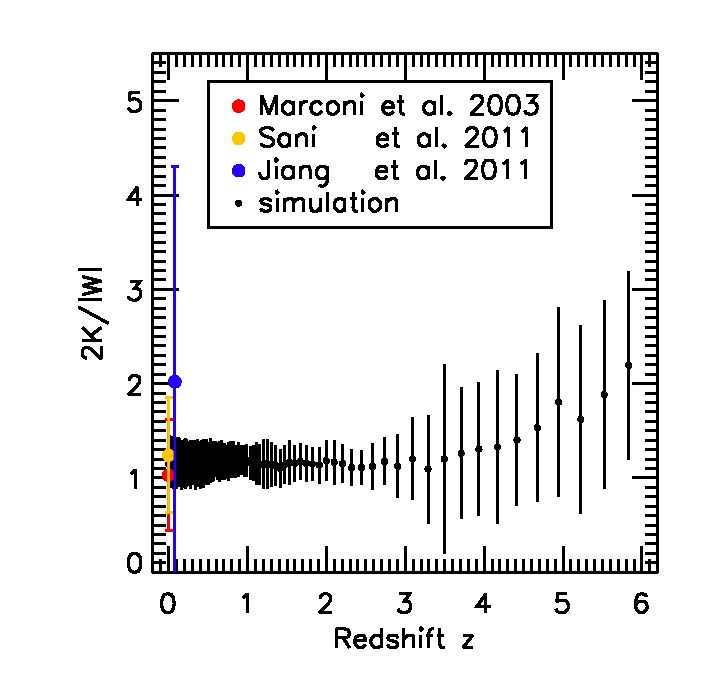}
\caption{\label{fig:vireq} The ratio of kinetic energy to the potential energy of stellar components within half-mass radius of the galaxy at different redshift from the Aquila Simulation. This ratio serves as an indicator of the dynamical state of a galaxy, for example, if the ratio is 1, it means the system is in virial equilibrium. The black error bar indicates the ratio range at each redshift, the black dot represents the mean value, the red dot is the value derived from observations of \cite{Marconi2003}, the yellow dot indicates data from \cite{Sani2011}, and blue dot indicates data from \cite{Jiang2011}. The samples of \cite{Marconi2003} and \cite{Sani2011} are dominated by massive elliptical galaxies which satisfy the local $\mbhsigma$ correlation, while that of \cite{Jiang2011} are low-mass systems with BHs of $10^5 - 10^6\, \Msun$, which show significant deviation from the local $\mbhsigma$ correlation.}
\end{center}
\end{figure}

We have shown in the previous section that the galaxies from our simulation fall on the the observed local $\mbhsigma$ relation at $z=0$, but there is a remarkable evolution in the relation with redshift, while observations of massive galaxies with spheroids show little evolution. In order to understand the origin of this relation and the cause of the evolution and discrepancy, we turn to the dynamical state of the galaxies as it determines the stellar velocity dispersion. 

A number of models have been proposed to explain the $\mbhsigma$ relation. \cite{Silk1998} assumed the formation of SMBH proceeds galaxies and the energetic wind from quasars expels the gas, resulting in a regulation of the growth of the spheroidal stellar component. \cite{Ciotti2007} argued that radiative feedback from BHs in ellipticals give rise to this scaling relation. \cite{Hopkins2007B} suggested that it takes a dynamical time for the feedback energy from the BHs to unbind the surrounding gas and reached a fundamental-plane, relating the BH mass with a combined parameter with stellar mass and velocity dispersion. 

The finding that the $\mbhsigma$ and the fundamental-plane relation are closely related suggest that galaxies satisfying these relations may have reached relaxation, or virial equilibrium (e.g., \citealt{Robertson2006, Hopkins2009A}), which satisfies the virial theorem:

\begin{equation}
\rm 2K+W=0 
\end{equation}
\noindent where $K = 3/2\Mstar \sigma^2$ is the kinetic energy of the stars, and $W = -3/5GM\Mstar/R_e$ is gravitational potential of the galaxy with mass $M$ within the effective radius $R_e$. 

To examine the dynamical state of our galaxies, we define a parameter virial ratio, $\lambda = \rm{2K /W}$. If the system is in virial equilibrium, then $\lambda =1$. Figure~\ref{fig:vireq} shows the resulting virial ratio $\lambda$ of the galaxy sample at different redshift from our simulation, in comparison with available observational data from \cite{Marconi2003},  \cite{Sani2011}, and \cite{Jiang2011}.  

\cite{Sani2011} studied the black hole-host galaxy scaling relations in Spitzer/IRAC band. The velocity  dispersion, K-band luminosity are taken from their Table 2 and the effective radius $R_e$ from their Table 3. The galaxy mass is then derived from the K-band luminosity with a mass-to-light ratio $\rm M/L = 1$. The number of total galaxies from this sample is 50.  In \cite{Marconi2003}, since the effective radius $R_e$ was measured in J band, we use the J band luminosity to compute stellar mass also assuming  $\rm M/L = 1$. This gave us total 37 galaxies. Note the samples of \cite{Marconi2003} and \cite{Sani2011} are dominated by massive elliptical galaxies which satisfy the local $\mbhsigma$ correlation, while that of \cite{Jiang2011} are low-mass systems with BHs of $10^5 - 10^6\, \Msun$, which fall below the local $\mbhsigma$ correlation. A sample of galaxies with low mass BHs is also included as we crosscheck the ones with both photometric measurements from \cite{Jiang2011} and velocity dispersion measurements from \cite{Xiao2011}. Luminosity of the bulge is estimated from the surface brightness in I band and the luminosity distance inferred from redshift z from \cite{Jiang2011}. The ratio $M/L$ in this band is also assumed to be 1. This gives us a sample of 30 galaxies. 

From Figure~\ref{fig:vireq}, a clear evolution of the virial ratio $\lambda$ is seen: the mean value flattens around $\sim 1.1 - 1.2$ up to $z\sim 2$, then increases gradually with redshift and reaches $\sim 2.2$ at $z \sim 6$. At $z=0$, our galaxies agree with the observed $\mbhsigma$ relation, and our mean value of the ratio lies in between those derived from observations by \cite{Marconi2003} and \cite{Sani2011}. This agreement is exciting, because it demonstrates that galaxies that fall on the $\mbhsigma$ correlation are indeed in virial equilibrium. On the other hand, most of the galaxies of \cite{Jiang2011} fall below the classical $\mbhsigma$ correlation, like ours at $z \gtrsim 3$, therefore, it is not surprising that its $\rm 2K/W$  ratio is larger than 1, similar to our result at $z \sim 6$. This means that the galaxies at high redshift in our simulation have not yet reached virial equilibrium.

This result suggests that the $\mbhsigma$ correlation may be a result of virial equilibrium of the galaxies. Qualitatively, from the criterion for virial equilibrium as  suggested by \cite{Binney2008}: 

\begin{equation}
<\sigma^{2}> \simeq 0.45\frac{GM}{R_{\rm e}}
\end{equation}

\noindent which gives a relation between the galaxy mass and the stellar velocity dispersion and effective radius:

\begin{equation}
M  \propto {\sigma}^2 R_{e}
\end{equation}
\noindent 

Since the luminosity of a galaxy, L ($\propto \Mstar$), is proportional to the stellar surface brightness and the projected surface area:
\begin{equation}
L  \propto {R_{\rm e}^2}I_{e} 
\end{equation}
\noindent  

Assuming a constant mass-to-light ratio $M/L$ and a constant surface brightness $I_e$, the above two relations lead to 
\begin{equation}
L(\propto \Mstar)  \propto {\sigma^4}
\end{equation}
\noindent 

From our results in the previous section, the self-regulated growth of stars and BHs in galaxies gives $\MBH \propto \Mstar$, therefore, 
\begin{equation}
 \MBH \propto \Mstar \propto {\sigma^4}
\end{equation}
\noindent 

This virial equilibrim-origin model can explain the discrepancy between observations and our simulation. In the observations, most of the galaxies, in particular massive ones and those with bulges, may have reached relaxation and virial equilibrium quickly, because the more massive a galaxy is, the shorter relaxation timescale it has \citep{Binney2008}. It is believed that merger can trigger a rapid growth of BHs and the build up of ellipticals and bulge component \citep{Springel2005A}, when large amount of shocked gas by gravitational interaction fuels BHs accretion and intense star formation \citep{Springel2005A}. This scenario is also supported by observations (e.g., \cite{Bennert2008, Alexander2005}). The spheroids formed through violent merger quickly relax and reach virial equilibrium. As a result, they satisfy the $\mbhsigma$ relation and show little or no evolution. On the other hand, low-mass systems take a longer time to reach virial equilibrium, thus they may show some evolution in the $\mbhsigma$.

\section{Discussions}

In order to achieve desired resolutions to resolve low-mass BH systems in cosmological simulations, we focus on a Milky Way-size halo in this paper. However, this has led to a serious limitation: the galaxy sample is very small. The large scattering due to the small number statistics in our results, in particular at high redshift, may have affected the actual evolution of the scaling relations we studied. 

In such a zoom-in simulation, there might be contamination by heavier low-resolution particles outside of the high-resolution zoom-in region \citep{Springel2008}. This could be more severe as we approach to the boundary of the zoom-in region. However, most of our galaxies are located well within the boundary. We checked the presence of low-resolution particles in all the subhalos which host BHs, and found that the mass fraction is very small, about $10^{-3} - 10^{-4}$. We conclude that the contamination is not significant in our galaxy sample.

The seeds of galactic BHs is an outstanding problem. It is unknown where the seed of the Milky Way came from. In our simulation, we follow the previous studies \cite{Li2007B, DiMatteo2008} and create seeds of $10^5\, \Msun$ in halos with mass above $10^{10}\, \Msun$. Although it is not our focus to reproduce exactly the same BH as in the Milky Way, our seeding scheme produced a $\sim 10^8\, \Msun$ BH in the main halo, much more massive than the current estimate of $10^6\, \Msun$ in the Milky Way. 

A wide range of seed masses, from $10 \, \Msun$ to $10^{6} \, \Msun$, has been proposed  \citep{Volonteri2010}. For examples, a small seed of $\sim 10^2\, \Msun$ by the collapsed remnants from the first stars \citep{Abel2002, Bromm2004, Gao2007, Yoshida2008}, or a massive one of $\sim 10^6\, \Msun$ by the catastrophic collapse of a supermassive star, or gas clump \citep{Carr1984, Bromm2003, Begelman2006, Tanaka2009}.

Recently, \cite{Volonteri2009} studied the effects of BH seeds on the BH scaling relations using semi-analytical merger trees. They investigated two seeding models, light seeds of $10^2\, \Msun$ from PopIII stars, and heavy seeds from direct gas collapse (which has a distribution function with a peak of  $10^5\, \Msun$), and suggested that the $\mbhsigma$ relation is a result of seeding mechanism and growth prescription, and that massive seeds produce better agreement with observations at $z=0$. They also predicted a large population of low-mass BHs at high redshifts, which has not been detected.  

We plan to improve the current study in future work by increasing the statistics. We will perform uniform cosmological hydrodynamic simulations in larger boxes in order to obtain a large number of galaxies at different redshift. We also plan to investigate different BH seeding schemes and the resulting BH -- host galaxy relations. 

\section{Summary}

We have investigated the $\mbhsigma$ and relations $\mbhhost$ relations in low-mass systems ($\MBH \sim \rm{10^{6}- 10^{8}}\, \Msun$) using the Aquila Simulation, a high-resolution cosmological hydrodynamic simulation focused on a Milky Way-size galaxy. The simulation included dark matter, gas dynamics, star formation, black hole growth, and feedback from both SNe and accreting BHs. Here is a list of our findings:

\begin{itemize}

\item The $\mbhsigma$ and $\mbhhost$ relations evolve differently with redshift: the former shows a remarkable evolution in both slope and normalization, while the latter shows little evolution in the redshift range $z=0-6$.

\item There is a close link between the $\mbhsigma$ relation and the dynamical state of the system -- the galaxies that fall on the observed correlation have a virial ratio $\lambda = \rm 2K/W$ close to 1, indicating that they are in virial equilibrium, while those that off the relation instead have a ratio larger than 1.

\item The star formation and black hole growth in galaxies are self-regulated -- the ratio between black hole accretion rate and star formation rate remains nearly constant over $z = 0-6$,  probably owing to feedback and gas availability in the galaxies.

\end{itemize}

These results suggest that observed scaling correlations have different origins: the $\mbhsigma$ relation may be the result of virial equilibrium, while the $\mbhhost$ relation may the result of self-regulated star formation and black hole growth in galaxies.

\acknowledgments

We thank Carlos Frenk for kindly providing the Aquila initial conditions to us, and Niel Brandt, Lars Hernquist, Volker Springel, Tiziana Di Matteo, Jonathan Trump, Yongquan Xue, Hide Yajima and Xinghai Zhao for stimulating discussions and helpful comments. Support from NSF grants AST-0965694 and AST-1009867 is gratefully acknowledged. We acknowledge the Research Computing and Cyberinfrastructure unit of Information Technology Services at The Pennsylvania State University for providing computational resources and services that have contributed to the research results reported in this paper (URL: http://rcc.its.psu.edu). The Institute for Gravitation and the Cosmos is supported by the Eberly College of Science and the Office of the Senior Vice President for Research at the Pennsylvania State University.

\end{document}